\documentclass{aa}  
\usepackage{graphicx}
\usepackage{natbib}
\bibpunct{(}{)}{;}{a}{}{,} 
\usepackage[varg]{txfonts}
\usepackage{float}
\usepackage{placeins}
\usepackage[backref,colorlinks,citecolor={blue},linkcolor={blue}]{hyperref}
\label{LastPage}
\begin{document} 

   \title{Dissecting stellar populations with manifold learning}

   \subtitle{I. Validation of the method on a synthetic Milky Way-like galaxy}

   \author{Neitzel, A. W.
          \inst{1, 2},
          Campante, T. L.
          \inst{1, 2},
          Bossini, D.
          \inst{3,4},
          Miglio, A.
          \inst{5, 6}
          }

   \institute{Instituto de Astrofísica e Ciências do Espaço, Universidade do Porto, CAUP,
              Rua das Estrelas, 4150-762 Porto, Portugal
              \\
              \email{andreaswneitzel@astro.up.pt}
        \and
             Departamento de Física e Astronomia, Faculdade de Ciências da Universidade do Porto, Rua do Campo Alegre, s/n, 4169-007 Porto, Portugal
        \and
            Dipartimento di Fisica e Astronomia Galileo Galilei, Università di Padova, Vicolo dell’Osservatorio 3, 35122 Padova, Italy
        \and
            INAF -- Osservatorio Astronomico di Padova, Vicolo dell’Osservatorio 5, 35122 Padova, Italy
        \and
            Department of Physics and Astronomy, University of Bologna, Via P.~Gobetti 93/2, 40129 Bologna, Italy
        \and
            INAF -- Osservatorio di Astrofisica e Scienza dello Spazio, Via P.~Gobetti 93/3, 40129 Bologna, Italy
             }

    \date{Received 30 July 2024 /
Accepted 21 January 2025}
             
 
  \abstract
   {Different stellar populations may be identified through differences in chemical, kinematic, and chronological properties, suggesting the interplay of various physical mechanisms that led to their origin and subsequent evolution. As such, the identification of stellar populations is key for gaining insight into the evolutionary history of the Milky Way galaxy. This task is complicated by the fact that stellar populations share significant overlap in their chrono-chemo-kinematic properties, hindering efforts to identify and define stellar populations.}
   {Our goal is to offer a novel and effective methodology that can provide deeper insight into the nonlinear and nonparametric properties of the multidimensional physical parameters that define stellar populations.}
   {For this purpose we explore the ability of manifold learning to differentiate stellar populations with minimal assumptions about their number and nature. Manifold learning is an unsupervised machine learning technique that seeks to intelligently identify and disentangle manifolds hidden within the input data. To test this method, we make use of \textit{Gaia} DR3-like synthetic stellar samples generated from the FIRE-2 cosmological simulations. These represent red-giant stars constrained by asteroseismic data from TESS.}
   {We reduce the 5-dimensional input chrono-chemo-kinematic parameter space into 2-dimensional latent space embeddings generated by manifold learning. We then study these embeddings to assess how accurately they represent the original data and whether they contain meaningful information that can be used to discern stellar populations.}
   {We conclude that manifold learning possesses promising abilities to differentiate stellar populations when considering realistic observational constraints.}


    \titlerunning{Dissecting stellar populations with manifold learning I}
    \authorrunning{Neitzel, A. W. et al.} 
   \maketitle

\nolinenumbers
\section{Introduction}
\label{sec:introduction}
    Galactic archaeology is the study of the assembly and evolution history of the Milky Way \citep{Freeman2002ARA&A..40..487F}. The Milky Way, as we observe it, is not a monolithic structure, but instead an agglomeration of different components that differ in structural, kinematic, and chemical properties \citep{Fuhrmann1998A&A...338..161F, Bensby2003A&A...410..527B, Reddy2006MNRAS.367.1329R, Bland-Hawthorn2016ARA&A..54..529B}. These include the thin and thick disk \citep{Gilmore1983MNRAS.202.1025G}, often identified as the low-$\alpha$ and high-$\alpha$ disk due to their dichotomy in [$\alpha$/Fe]--[Fe/H] space \citep{Adibekyan2012A&A...545A..32A}, the stellar halo \citep{Helmi2008A&ARv..15..145H} and the bulge \citep{Shen2016ASSL..418..233S}. Other prominent stellar populations include stellar populations resulting from accretion events, such as streams from disrupted galaxies \citep[e.g., Saggitarius Dwarf Spheroidal galaxy;][]{Ibata1994Natur.370..194I, Vasiliev2020MNRAS.497.4162V} and the famous \textit{Gaia}-Enceladus \citep{Helmi2018Natur.563...85H}, considered the last large merger event in the history of the Galaxy.

    In addition to their differences in chemical and kinematic properties, dating these events is key. Stellar ages allow us to better constrain the timeline of accreted events \citep{Chaplin2020NatAs...4..382C, Montalban2021NatAs...5..640M} and the formation of different stellar populations \citep{SilvaAguirre2018MNRAS.475.5487S}. The current state of the art in stellar age estimation comes from asteroseismology, i.e., the study and observation of stellar oscillations \citep[see, e.g.,][]{Garcia2019LRSP...16....4G}. When global asteroseismic parameters are used, uncertainties in the age estimations of hydrogen-shell burning (RGB) and helium-core burning (HeCB) red-giant stars can reach up to 20--25\% \citep{SilvaAguirre2020ApJ...889L..34S, Mackereth2021MNRAS.502.1947M, Stello2022MNRAS.512.1677S, Campante2023AJ....165..214C}. When the signal-to-noise ratio (SNR) and observation duration are sufficient, individual modes can also be used to double the precision in age \citep{Metcalfe2014ApJS..214...27M}. This has been achieved in previous asteroseismic missions and is an expected goal of the upcoming Planetary Transits and Oscillations of Stars mission \citep[PLATO;][]{Miglio2017AN....338..644M}. By expanding our understanding of the Milky Way, we can impose more precise constraints on stellar nucleosynthesis and the timescales of formation of the various Galactic components \citep{Matteucci2021A&ARv..29....5M}, and develop gas infall models that can provide a theoretical framework for the evolution of the Galaxy \citep{Noguchi2018Natur.559..585N, Spitoni2019A&A...623A..60S, Spitoni2023A&A...670A.109S}.

    Scientific advances in Galactic archaeology have greatly benefited from large-scale surveys, particularly those including asteroseismic parameters \citep{Anders2017A&A...597A..30A, Miglio2021A&A...645A..85M}. The latest \textit{Gaia} Data Release 3 \citep[DR3;][]{Lindegren2021A&A...649A...2L, Gaia2023A&A...674A...1G} provides astrometry and broadband photometry for $1.8 \times 10^9$ objects. Of these, $3.38 \times 10^7$ contain radial velocities, providing a complete 6-dimensional picture of stellar kinematics. Spectroscopic surveys, such as Galactic Archaeology with High Efficiency and Resolution Multi-Element Spectrograph \citep[GALAH;][]{Buder2021MNRAS.506..150B}, Large Sky Area Multi-Object Fiber Spectroscopic Telescope \citep[LAMOST;][]{Cui2012RAA....12.1197C}, and Apache Point Observatory Galactic Evolution Experiment \citep[APOGEE;][]{Majewski2017AJ....154...94M}, provide chemical abundances, with APOGEE DR17 providing spectroscopic measurements for 372,458 targets \citep{Abdurro'uf2022ApJS..259...35A}. Future surveys collect from the 4-metre Multi-Object Spectroscopic Telescope \citep[4MOST;][]{deJong2019Msngr.175....3D}, and the William Herschel Telescope (WHT) Enhanced Area Velocity Explorer \citep[WEAVE;][]{Dalton2014SPIE.9147E..0LD, Jin2024MNRAS.530.2688J}, are expected to significantly expand the volume of available data. Asteroseismic parameters have been obtained from surveys such as the Convection, Rotation and planetary Transits mission \citep[CoRoT;][]{Baglin2009IAUS..253...71B}, the \textit{Kepler} and K2 missions \citep{Borucki2010Sci...327..977B, Howell2014PASP..126..398H}, and Transiting Exoplanets Survey Satellite \citep[TESS;][]{Ricker2015JATIS...1a4003R}, with the PLATO mission scheduled for launch in 2026 \citep{Rauer2014ExA....38..249R}. The chronological, chemical, and kinematic data obtained from these surveys can be cross-matched to create stellar catalogs with complete chrono-chemo-kinematic parameters, which are expected to provide insight into the stellar populations of the Galaxy.

    As the volume of data increases and new variables are included, the complexity of extracting valuable information from the data also increases. Insight into stellar populations may be hidden in the form of complex, nonlinear relations that span the multidimensional data. Such relations become harder to uncover as the number of parameters (i.e., dimensions) increases. A potential solution lies in data-driven approaches such as machine learning. Machine learning (ML) is the field of computer programs that are capable of autonomously solving complex problems without following explicit instructions. This is accomplished through statistical models used to draw inferences from patterns in the data. ML methodologies are gaining popularity in Galactic archaeology, with methods frequently tested in $\Lambda$CDM \citep{Davis1985ApJ...292..371D} cosmological zoom-in simulations, such as FIREbox \citep{Feldmann2023MNRAS.522.3831F} and Auriga \citep{Grand10.1093/mnras/stae1598}. \cite{Ostdiek2020A&A...636A..75O} trained deep learning neural networks on synthetic stars from the FIRE-2 simulations \citep{Hopkins2018MNRAS.480..800H} to develop a catalog of accreted stars in the Milky Way. \cite{Necib2020ApJ...903...25N} applied Gaussian Mixture Models (GMM) and Density-Based Spatial Clustering of Applications with Noise \citep[DBSCAN;][]{Ester1996ADA} to the aforementioned catalog in the search for accreted structures. In \cite{Nikakhtar2021ApJ...921..106N} and \cite{Stokholm2023MNRAS.524.1634S}, GMM was used to search for stellar populations. \cite{Sante10.1093/mnras/stae1398}, artificial neural networks, decision trees and dimensionality reduction techniques were applied on synthetic Milky Way-like galaxies from the ARTEMIS simulation \citep{Font2020MNRAS.498.1765F} to train a model capable of differentiating accreted from in situ stars. In \cite{Queiroz2023A&A...673A.155Q}, the Bayesian isochrone-fitting code \texttt{StarHorse} was used to derive stellar ages, which, in conjunction with chemical abundances, permitted the construction of a chrono-chemical stellar sample that was then analyzed using the t-distributed Stochastic Neighbor Embedding \citep[t-SNE;][]{vanDerMaaten2013} dimensionality reduction technique and the Hierarchical DBSCAN \citep[HDBSCAN;][]{Campello2013-lh} clustering algorithm in the search for stellar populations.
    
    In this paper, we tackle the question of how to effectively uncover patterns and commonalities in the data, as well as 'disentangle' stellar populations embedded in high-dimensional data, through a technique referred to as manifold learning. For this purpose we make use of the Uniform Manifold Approximation and Projection for Dimension Reduction algorithm \citep[UMAP;][]{McInnes2018arXiv180203426M}, a state-of-the-art non-parametric manifold learning dimensionality reduction algorithm. We explore the potential of UMAP to assist in identifying stellar populations while making minimum assumptions about the nature of the data. To validate the method, we test it on four different synthetic stellar samples with known ground truth.

    The structure of this work is as follows. Section \ref{sec:data} describes the construction of the synthetic stellar samples. Section \ref{sec:methods} provides a brief review of the inner workings of UMAP and a detailed explanation of its application to the test samples. The results are presented in Sect. \ref{sec:results}. Finally, we provide a summary of the results and our conclusions in Sect. \ref{sec:summary_conclusions}. Further plots are provided in the \href{https://doi.org/10.5281/zenodo.14749285}{Appendix}.

\section{Data}
\label{sec:data}

\subsection{Synthetic Milky Way-like galaxy}
\label{sec:synth_galaxy}
    Feedback In Realistic Environments \citep[FIRE-1, later FIRE-2;][]{Hopkins2014MNRAS.445..581H, Hopkins2018MNRAS.480..800H} is a numerical implementation of feedback mechanisms that regulate star formation in the \texttt{GIZMO} cosmological evolution code \citep{Hopkins2015MNRAS.450...53H}. The \textit{Latte} project \citep{Wetzel2016ApJ...827L..23W, Hopkins2018MNRAS.480..800H} aims to replicate Milky Way-like galaxies in the FIRE-2 cosmological zoom-in simulations, resulting in three synthetic galaxies with Milky Way-like mass and morphology (\texttt{m12i}, \texttt{m12f} and \texttt{m12m}). For each of these galaxies, \cite{Sanderson2020ApJS..246....6S} selected three heliocentric-like local standards of rest (LSR) and implemented the \texttt{Ananke} framework, which decomposes the original stellar particles of the FIRE-2 simulation at redshift $z = 0$ into individual stars and subsequently convolves their observable properties using \textit{Gaia} DR2 uncertainties, thereby generating synthetic \textit{Gaia} DR2-like surveys from the FIRE-2 simulations. This procedure was later updated with \textit{Gaia} DR3 uncertainties in \cite{Nguyen2024ApJ...966..108N}. The full FIRE-2 public data release is provided in \cite{Wetzel2023ApJS..265...44W}. For this work, we use the \texttt{m12f-lsr0} \textit{Gaia} DR3-like synthetic catalog.

\begin{figure}
    \centering
    \includegraphics[width=\columnwidth]{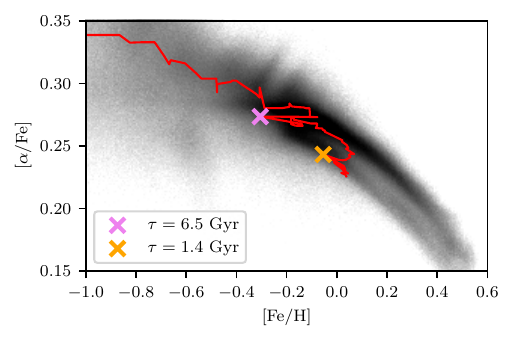}
    \caption{Chemical abundance scatter plot for the redshift $z = 0$ snapshot of the \texttt{m12f} synthetic galaxy. The red curve traces the average $\mathrm{[\alpha/Fe]}$ and $\mathrm{[Fe/H]}$ positions in equally spaced intervals of $\Delta \tau = 100 ~\mathrm{Myr}$. The tendency is to evolve from the top left to the bottom right, i.e., from high $\mathrm{[\alpha/Fe]}$ and low $\mathrm{[Fe/H]}$ to low $\mathrm{[\alpha/Fe]}$ and high $\mathrm{[Fe/H]}$.}
    \label{fig:m12f_chemical_evolution}
    \end{figure}

\begin{figure*}
    \centering
    \includegraphics[width=\textwidth]{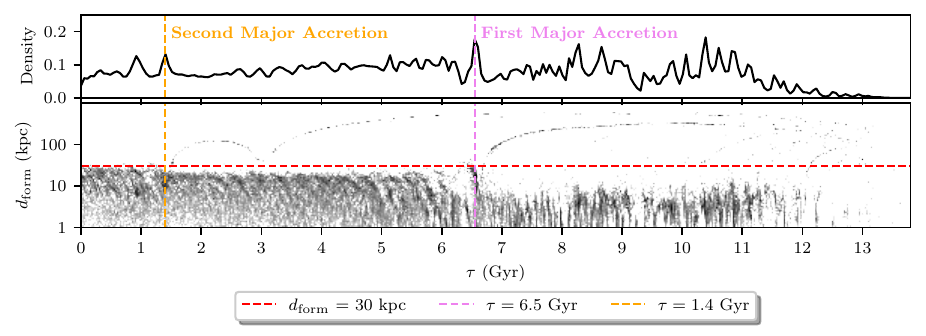}
    \caption{Redshift $z = 0$ snapshot of the \texttt{m12f} synthetic galaxy considering only stellar particles located within $30 ~\mathrm{kpc}$ from the galactic center at that particular instant in time. \textbf{Top:} Gaussian kernel density estimate for the age distribution of the \texttt{m12f} stellar particles. Spikes are noticeable at ages $6.5$ and $1.4 ~\mathrm{Gyr}$. \textbf{Bottom:} Scatter plot of $d_\mathrm{form}$ as a function of age. Accreted galaxies are shown as curves of continuously generating stellar particles approaching from distances $d > 100 ~\mathrm{kpc}$, eventually colliding with \texttt{m12f}. These collisions coincide with spikes in star formation.}
    \label{fig:m12f_dform}
    \end{figure*}

    The properties of the synthetic galaxy \texttt{m12f} are described in \cite{Hopkins2018MNRAS.480..800H} and summarized below. It is a spiral galaxy with virial mass $M^\mathrm{vir} = 1.6 \times 10^{12} ~\mathrm{M_\odot}$ (virial radius $R^\mathrm{vir} = 306 ~\mathrm{kpc}$) and a stellar mass $M = 8.0 \times 10^{10} ~\mathrm{M_\odot}$. These values are comparable to those of the Milky Way, for which $M^\mathrm{vir} \approx 10^{12} ~\mathrm{M_\odot}$ and $M \approx 5 \times 10^{10} ~\mathrm{M_\odot}$ \citep{Bland-Hawthorn2016ARA&A..54..529B}. In the cosmological simulation, the first $5.8 ~\mathrm{Gyr}$ ($8 < \tau < 13.8 ~\mathrm{Gyr}$) of \texttt{m12f} were characterized by an initial period of gas infall and several minor accretion events into its primordial dark-matter halo. This was followed by a brief quiescent period ($6.5 < \tau < 8 ~\mathrm{Gyr}$), which allowed the galaxy to begin forming the disk before experiencing a major collision with another galaxy ($\tau \approx 6.5 ~\mathrm{Gyr}$). This event is dubbed the ``first major accretion event''. Around $5 ~\mathrm{Gyr}$ later, \texttt{m12f} experienced what we call its ``second major accretion event'' ($\tau \approx 1.4 ~\mathrm{Gyr}$). Both these major accretion events contributed metal-poor gas, which mixed with the in situ gas, leading to episodes of elevated star formation with distinct chemical signatures. This is due to the different chemical enrichment histories of their progenitor nebulae, consistent expectations following the \textit{Gaia}-Enceladus merger \citep[see e.g.,][]{Vincenzo2019MNRAS.487L..47V, Giribaldi2023A&A...673A..18G, Ciuca2024MNRAS.528L.122C}. This effect is shown in Fig. \ref{fig:m12f_chemical_evolution} and confirmed in Fig. \ref{fig:m12f_dform}, where the distance of formation ($d_\mathrm{form}$) of stellar particles is plotted. The rebound at $\tau \approx 3 ~\mathrm{Gyr}$ in Fig. \ref{fig:m12f_dform} does not constitute an accretion event, but rather a flyby of the galaxy in which it ventured close to \texttt{m12f}, though not close enough to cause a collision. Following the prescription of \cite{Nikakhtar2021ApJ...921..106N}, we define stars with $d_\mathrm{form} > 30 ~\mathrm{kpc}$ as accreted. We also observe a very narrow chemical dichotomy at high [Fe/H] with a $\Delta \mathrm{[\alpha/Fe]}$ separation of about $0.02 ~\mathrm{dex}$ (see Fig.~\ref{fig:m12f_chemical_evolution}). This is much smaller than the chemical separation between the Milky Way's low-$\alpha$ and high-$\alpha$ disks, which is about $0.2 ~\mathrm{dex}$ \citep{Vincenzo2021MNRAS.508.5903V}.

    The local standard of rest \texttt{m12f-lsr0} is located at a vertical distance from the galactic plane of $Z_\mathrm{Gal} = 0 ~\mathrm{pc}$ and at a distance $R_\mathrm{Gal} = 8.2 ~\mathrm{kpc}$ from the galactic center, similar to estimates of the solar position relative to the center of the Milky Way \citep[for a comprehensive literature review, see][chap.~3.2]{Bland-Hawthorn2016ARA&A..54..529B}. Its azimuthal velocity around the galactic center is $V_\mathrm{LSR} = 226 ~\mathrm{km\,s^{-1}}$, consistent with its neighborhood and comparable to the azimuthal velocity of the solar system's local standard of rest in the Milky Way \citep[see][chaps.~3.4 and 6.4]{Bland-Hawthorn2016ARA&A..54..529B}. The \texttt{Ananke} framework simulates a \textit{Gaia} DR3-like survey, accounting for extinction from gas and the uncertainties of \textit{Gaia} DR3, resulting in $4,265,816,043$ observable stars, of which $138,536,719$ have radial velocities \citep[for more details, see][]{Nguyen2024ApJ...966..108N}. Rather than using the entire survey, we will instead construct realistic observable samples, both in terms of the number of stars and selection function, as described in Sect.~\ref{sec:synth_samples}.

We now define the 5-dimensional input chrono-chemo-kinematic parameter space that will be adopted in this study, which includes the following stellar parameters:
    \begin{itemize}
        \item The stellar age, $\tau$, in Gyr.
        \item Metallicity, [Fe/H].
        \item The abundance ratio of $\alpha$-process elements over iron, [$\alpha$/Fe], which is defined in FIRE-2 as [Mg/Fe].
        \item The azimuthal velocity of a star around the galactic center, $V$, in $\mathrm{km\,s^{-1}}$.
        \item The orthogonal cylindrical velocity components to the azimuthal velocity, $U$ and $W$, combined into $\sqrt{U^2 + W^2}$, in $\mathrm{km\,s^{-1}}$.
    \end{itemize}
    The stellar age and abundance ratios are provided by the simulations. We use [Fe/H] and [$\alpha$/Fe], as these respectively trace type Ia supernovae (Ia SNe) and core-collapse supernovae (CC-SNe), serving as important galactic clocks \citep{Arcones2023A&ARv..31....1A}. We do not consider the remaining elemental abundances available in the simulation. The seismic stellar samples (samples B, C and D; see Sect.~\ref{sec:synth_samples}) contain cool stars that do not possess sufficiently hot photospheric temperatures to excite helium. The surface [C/N] ratio is altered by processes occurring during stellar evolution (e.g., the first dredge-up) and hence does not qualify as a galactic clock. The remaining elements are $\alpha$-process elements, whose abundance ratios are highly (linearly) correlated with [Mg/Fe] (Pearson correlation coefficient $\rho > 0.8$). For this reason, we opted to keep [Mg/Fe] as the only proxy for [$\alpha$/Fe]. Some FIRE-2 simulations do include r-process elements \citep{Voort2015MNRAS.447..140V}, but this is not the case for the Milky Way-like galaxies present in the \textit{Latte} suite. Furthermore, to calculate the stellar velocities in a left-handed cylindrical coordinate system, \{$U$, $V$, $W$\}, we make use of the \texttt{Astropy} \citep{Astropy2022ApJ...935..167A} package, using as input parameters the parallax, angular position, radial velocity and proper motions of the stars, as obtained from the \texttt{Ananke} \textit{Gaia} DR3-like survey. The LSR is set at a distance of $R_\mathrm{Gal} = 8.2 ~\mathrm{kpc}$ from the galactic center and perfectly centered in the galactic plane ($Z_\mathrm{Gal} = 0 ~\mathrm{pc}$).

\subsection{Synthetic stellar samples}
\label{sec:synth_samples}
We generated four different \textit{Gaia} DR3-like synthetic stellar samples from the \texttt{m12f-lsr0} data set: samples A, B, C and D. Sample A will serve as the ground truth reference, being the most diverse sample, with 75,000 randomly selected stars in all evolutionary states within $3 ~\mathrm{kpc}$ of the LSR observer. We select stars that have available radial velocities, that lie in the apparent \textit{Gaia}-band magnitude range $6 < G < 13$, and further match the \textit{Gaia} precision criteria in \cite{Gaia2018A&A...616A..10G}:
    \begin{itemize}
        \item Fractional parallax uncertainty $< 0.1$.
        \item Uncertainty in the \textit{Gaia} $G$ band $< 0.022 ~\mathrm{mag}$.
        \item Uncertainty in the \textit{Gaia} $G_\mathrm{BP}$ band $< 0.054 ~\mathrm{mag}$.
        \item Uncertainty in the \textit{Gaia} $G_\mathrm{RP}$ band $< 0.054 ~\mathrm{mag}$.
        \item Extinction $A_G < 0.015 ~\mathrm{mag}$.
    \end{itemize}

The remaining stellar samples were chosen to mimic red-giant stars, constrained by asteroseismic data from TESS. Red-giant stars are ubiquitous, visible at large distances due to their high luminosities, and they further possess deep convective envelopes that allow high-SNR detection of oscillation modes. These properties make them ideal targets for probing the Galactic structure across space and time. The TESS mission enabled the creation of a seismic red-giant star catalog containing the frequency of maximum oscillation power ($\nu_\mathrm{max}$) and asteroseismic mass estimates for $158,505$ red giants, spanning a large sky area \citep{Hon2021ApJ...919..131H}. Following the recommendations in \citet{Hon2021ApJ...919..131H}, we preserve only the catalog stars with \texttt{mass\_flag = 1} and \texttt{ruwe} $\leq 1.40$, where the former condition implies $0.6 \leq M \leq 2.9 ~\mathrm{M_\odot}$, and the latter indicates an adopted astrometric solution likely to correspond to a single star. We further select stars with relative uncertainties below $10\%$ in effective temperature ($T_\mathrm{eff}$), luminosity ($L$), distance ($d$), and the \textit{Gaia}-band magnitude ($G$). This left us with $137,404$ seismic red-giant stars spanning almost the entire sky and covering a distance distribution with a mode of $800 ~\mathrm{pc}$.

\begin{figure}
    \centering
    \includegraphics[width=\columnwidth]{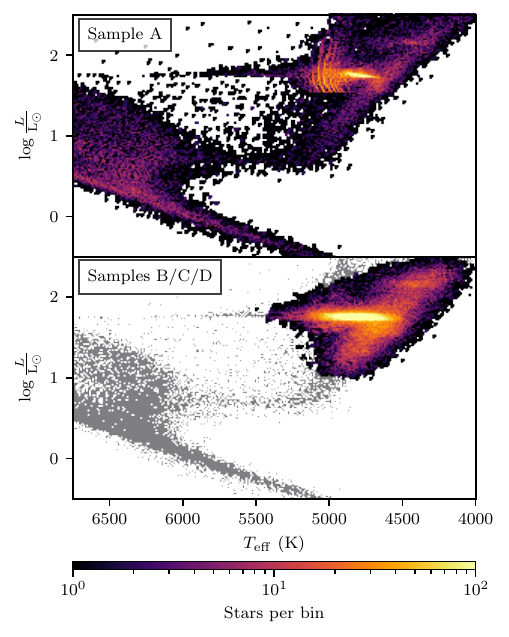}
    \caption{HR diagram of the synthetic stellar samples showing the main sequence, subgiant phase, red-giant branch and the red clump. \textbf{Top:} Sample A. \textbf{Bottom:} Samples B, C and D (color-coded) overlapping the sample A stars (in gray).}
    \label{fig:HRD_A_vs_BCD}
    \end{figure}

To serve as a ground truth TESS-like sample, we generated sample B. Besides the same requirements as for sample A, the selection function of sample B is chosen to mimic the evolutionary states and physical parameters of the TESS seismic red-giant catalog. Firstly, to blur the otherwise discrete evolutionary track gridpoints in the Hertzsprung--Russell (HR) diagram, the $T_\mathrm{eff}$ and $L$ values are perturbed according to Gaussian distributions with mean, $\mu$, equal to their true value and standard deviation, $\sigma$, equal to the median relative uncertainties found in the TESS seismic red-giant catalog ($0.02$ for $T_\mathrm{eff}$ and $0.04$ for $L$). Secondly, we replicated the TESS selection function by using kernel density estimators (KDE) and k-dimensional (KD) tree nearest neighbor algorithms. For the stellar mass, $M$, we constructed a 1-dimensional KDE and sampled stars to approximate the seismic TESS catalog's mass distribution. We then wanted to perform independent 2-dimensional samplings of the $T_\mathrm{eff}$--$\nu_\mathrm{max}$ diagram, the right-ascension and declination sky coverage, $\mathrm{RA}$--$\mathrm{DEC}$, and of the distance from the observer with the \textit{Gaia} band, $d$--$G$. For computational reasons, we forego the use of KDEs in the 2-dimensional case, instead opting for the faster KD tree algorithm. This algorithm uses binary space partitions to efficiently find nearest neighbors of query points, allowing us to sample points within a volume of interest. We considered the sky position in the crossmatch because TESS observations are biased towards its southern and northern continuous viewing zones and the fact that TESS misses observations for parts of the sky. The combination of $T_\mathrm{eff}$, $\nu_\mathrm{max}$ (calculated via the scaling relation provided in \citealt{Stello2008ApJ...674L..53S}) and $M$ ensures that the appropriate evolutionary states are selected. We discard a crossmatch in luminosity to avoid introducing bias due to reddening effects. The resulting crossmatch contains $76,168$ stars, about half the size of the reference catalog. Figure \ref{fig:HRD_A_vs_BCD} compares the parameter spaces of samples A and B, with the latter being shared with samples C and D, which we will now go on to describe. We provide a comparison of samples A, B and the TESS seismic red-giant catalog in Fig. \href{https://doi.org/10.5281/zenodo.14749285}{B.1} of the Appendix.
    
Finally, we generated samples C and D from sample B such that their chemical abundances and stellar ages are perturbed from their true values, with the kinematics being adopted from the error-convolved \textit{Gaia} DR3-like values. The differences between samples C and D are as follows. Sample C will contain stellar age and chemical abundance perturbations based on realistic observational uncertainties, whereas sample D will have enhanced-precision (i.e., optimistic) perturbations. Firstly, we calculate the stellar velocities in a left-handed cylindrical coordinate system, \{$U$, $V$, $W$\}, using the error-convolved parallax, angular position, radial velocity, and proper motions. These kinematics are shared by both samples. Secondly, we perturb the values for age by $20\%$ for sample C (in line with typical asteroseismic stellar dating estimates for red-giant stars) and by $10\%$ for sample D. Finally, we perturb the values of $\mathrm{[Fe/H]}$ and $\mathrm{[\alpha/Fe]}$ according to the median internal uncertainties of APOGEE DR17 for sample C ($0.0095$ and $0.0151 ~\mathrm{dex}$, respectively) and the 25th percentile uncertainties for sample D ($0.0076$ and $0.0123 ~\mathrm{dex}$, respectively). We note that a conservative uncertainty on the chemical abundances would be on the order of $0.1  ~\mathrm{dex}$, hence larger than the $\sim0.01  ~\mathrm{dex}$ adopted here. This reasoning comes from the fact that the chemical separation between the high-$\alpha$ and low-$\alpha$ disks is significantly smaller for \texttt{m12f} than observed for the Milky Way (see Fig.~\ref{fig:m12f_chemical_evolution}). As a result, a perturbation of $0.1 ~\mathrm{dex}$ would cause a level of mixing in chemical space that would make the disks indistinguishable. We provide a comparison of samples B, C, and D in Fig.~\href{https://doi.org/10.5281/zenodo.14749285}{B.2} of the Appendix.

\section{Methods}
\label{sec:methods}
    The identification and differentiation of stellar populations is a problem of stellar group membership assignment that considers commonalities in physical parameter space and plausible theories to explain their origin. Clustering algorithms such as K-Means \citep{macqueen:1967, AbiodunIKOTUN2023178}, DBSCAN \citep{Ester1996ADA}, and HDBSCAN \citep{Campello2013-lh}, as well as generative models such as Gaussian Mixture Models \citep[GMM;][]{Reynolds2009}, may provide a first step towards the assignment of group membership, however, it is crucial to consider their limitations. Generative model algorithms like GMM make parametric assumptions on the distributions that generate the data, which may pose a problem if the data are distributed according to a different parametric distribution or if the data do not follow any parametric distribution. Clustering algorithms do not make parametric assumptions about the data. However, there are nuances which may lead to some of them being more appropriate than others depending on the circumstances. Even when taking such nuances into account, the validity of the resulting clusters may be difficult to assess, particularly for large datasets where computational memory becomes a limiting factor. Furthermore, it can be shown that applying clustering algorithms to high-dimensional data may be suboptimal compared to clustering in lower-dimensional spaces, provided these lower-dimensional spaces are constructed to emphasize potentially meaningful local structure in the data. An example using HDBSCAN is provided in Appendix \href{https://doi.org/10.5281/zenodo.14749285}{D}. Given these considerations, we were motivated to pursue a strategy of dimensionality reduction.
    
    Dimensionality reduction is the transformation of high-dimensional input data into a lower-dimensional output space. A popular, yet limited, unsupervised learning algorithm used for this purpose is Principal Component Analysis \citep[PCA;][]{Pearson01111901, Francis1999ASPC..162..363F}, which identifies the principal components of the input dataset, defined as the eigenvectors of the covariance matrix. An $N$-dimensional input space results in $N$ principal components, from which the top $n < N$ are selected in order to reduce the dimensionality of the data. This method, popular due to its speed and deterministic nature, is limited to linear transformations of the data. Furthermore, feature importance in PCA is defined by a global measure of variance, which is sensitive to noise and fails to capture structural nuance at a local level. For complex nonlinear cases, a more general approach is required.

    Our methodology of choice must be capable of generalizing to non-linear and non-parametrically distributed data, in a manner that allows us to confidently assess the validity of the clusters found. One such algorithm is t-SNE \citep{vanDerMaaten2013}, which works by determining the neighborhood relationships of the high-dimensional input data and then projecting them into a low-dimensional output space, known as a latent space embedding. This method is well-suited for nonlinear, nonparametrically distributed datasets and provides a visualization that helps make more confident decisions on how to segregate the data. It is important to note that the projection through the construction of a latent space embedding operates differently from standard projections, such as PCA. Whereas PCA is a deterministic algorithm limited to linear transformations that uses global variance as a metric for feature importance, latent space embedding iteratively evolves the dimensionally reduced space by prioritizing local neighborhood preservation. It is further capable of handling local non-linear relationships, emphasizing the presence of complex data relationships that can then be more easily clustered. This is the strategy we ultimately adopt, though using a different algorithm: UMAP.
    
    We justify the choice of UMAP in the following subsection. A comparison of UMAP's effectiveness with PCA and t-SNE is provided in Appendix \href{https://doi.org/10.5281/zenodo.14749285}{D}.
    
\subsection{UMAP}

    The Uniform Manifold Approximation and Projection for Dimension Reduction \citep[UMAP;][]{McInnes2018arXiv180203426M} algorithm is a method that combines aspects of manifold learning and topological data analysis for dimensionality reduction. Firstly, UMAP constructs a fuzzy topological representation of the input high-dimensional data, wherein data points are connected by weighted edges based on a similarity criterion that is quantified using a probability distribution that measures the likelihood of points being connected. Secondly, UMAP captures neighborhood relationships using a local approach to preserve the local structure of the data. Finally, a lower-dimensional latent space embedding is constructed by minimizing a cost function that balances the preservation of local neighborhood relationships and the overall global structure. This low-dimensional embedding is subsequently iteratively adjusted to achieve a configuration that best aligns with the underlying manifold. The final output is an embedding that contains information on both the local and global patterns and commonalities found in the data, depending on the choice of hyperparameters. The most important hyperparameters are \texttt{n\_neighbors} and \texttt{min\_dist}. The former constrains the size of the local neighborhood analyzed by UMAP when attempting to learn the manifold structure of the data, and the latter controls how tightly UMAP aggregates points together in the latent space embedding. Tuning these hyperparameters will determine whether the resulting embedding emphasizes local or global structure. An exaggerated emphasis on the finer structure can lead to noisy results, whereas emphasis on macroscopic structure can instead result in an oversmoothed, non-descriptive embedding. For these reasons, it is important to choose these hyperparameters carefully to balance out these effects.

    In this paper, we applied the UMAP algorithm using the \texttt{Python} \texttt{UMAP} library on the synthetic samples A, B, C and D generated in Sect. \ref{sec:synth_samples}. The input was input a 5-dimensional chrono-chemo-kinematic parameter space spanning the variables $\{V, UW, \mathrm{[Fe/H]}, \mathrm{[\alpha/Fe]}, \tau \}$, where $UW = \sqrt{U^2 + W^2}$. Since these variables span significantly different ranges of magnitudes, which may influence the algorithm's output, each variable is scaled using the \texttt{StandardScaler} function in \texttt{Sklearn} \citep{pedregosa2011scikit}. Our choice of UMAP hyperparameters aimed to balance local structure with global structure. For this purpose, we varied \texttt{min\_dist} and \texttt{n\_neighbors} for sample A and visually inspected the resulting embedding, settling in the end on $\texttt{min\_dist} = 0$ and $\texttt{n\_neighbors} = 15$. We found that meaningful embeddings were obtained after $1000$ epochs. However, we conservatively set the number of epochs to $10,000$ in all UMAP runs to ensure that the result has adequately approximated its final form. For simplicity, we use an Euclidean metric. Moreover, we reduce the 5D input data to a 2D representation for ease of visual inspection. An objective choice of the number of dimensions for the latent space embedding could be explored by calculating the intrinsic dimension, especially for input parameter spaces with a higher number of dimensions. Since ours has only 5 dimensions, we do not perform any such measure of intrinsic dimensionality.

\subsection{Trustworthiness}

    An objective measure of the validity of an embedding is the trustworthiness score, $T$ \citep{Venna10.1007/3-540-44668-0_68}, which depends on the number of nearest neighbors, $k$. Note that $k$ should not be confused with the \texttt{n\_neighbors} hyperparameter. The trustworthiness score validates how closely the low-dimension embedding resembles the high-dimension input in terms of the preservation of neighborhood. In other words, an embedding is trustworthy if stars that are neighbors in the input parameter space remain neighbors in the embedding. The trustworthiness score is defined as:
    \begin{align}
        T(k) = 1 - \frac{2}{Nk(2N - 3k - 1)}
        \sum_{i=1}^N \sum_{\vec{x}_j \in U_k (\vec{x}_i )}
        \left( r(\vec{x}_i, \vec{x}_j) - k \right) \, ,
        \label{eqn:trustworthiness}
    \end{align}
    where $N$ is the number of stars, $\vec{x}_i$ is the coordinate of the star $i$, and $U_k (\vec{x}_i)$ is the set of $k$ nearest neighbors of $\vec{x}_i$ in the embedding space that are not part of the $k$ nearest neighbors of $\vec{x}_i$ in the input space. The term $r(\vec{x}_i, \vec{x}_j) \in \mathbb{N}$ ranks the proximity of each coordinate $\vec{x}_j \in U_k (\vec{x}_i)$ relative to the coordinate of the star, $\vec{x}_i$, as measured in the input space. Thus, $r(\vec{x}_i, \vec{x}_j) > k \, \forall \, \vec{x}_j \in U_k (\vec{x}_i)$, meaning the term in the summation will be positive. If the $k$ nearest neighbors of $x_i$ in the embedding space are all within the $k$ nearest neighbors of $x_i$ in the input space, then the set $U_k (\vec{x}_i)$ is empty for that $i$ and the step is skipped.
    
    Equation (\ref{eqn:trustworthiness}) consists of two terms. The first term represents a perfect score of 1, while the second term is a penalty term, checking whether the neighborhood has been preserved for each star in the sample. If it has not, then it will decrease the score in proportion to the severity of the discrepancy, as determined by the rank term $r(\vec{x}_i, \vec{x}_j)$. This second term is normalized to constrain the trustworthiness score in the range $T(k) \in [0, 1]$, with values near 0 indicating heavy penalization. We calculate $r(\vec{x}_i, \vec{x}_j)$ using the same metric as was used in UMAP. Since $T$ varies as a function of the number of nearest neighbors, $k$, similar to the UMAP hyperparameter \texttt{n\_neighbors}, we set $k = \texttt{n\_neighbors}$. We adopt the conservative heuristic that an embedding is considered trustworthy if the trustworthiness score is $T(k=\texttt{n\_neighbors}) > 0.90$.

    Once the embedding passes the trustworthiness test, the next step is to select the learned structures. We look for evidence of differentiated structures, by which we mean topographically interesting components that can be clearly distinguished from one another. For example, internally coherent components that are disjoint from one another may indicate different stellar populations with significantly different physical parameters. Conversely, an embedding is considered uninformative if it lacks topographically interesting features. Uninteresting features may include spherical shapes and homogeneity. Since these structures can vary in shape and size, we use an unsupervised clustering algorithm to objectively identify and select structures.

\subsection{HDBSCAN}

    Though several unsupervised clustering algorithms exist, all of them possess their own specific limitations. To give two examples, the popular K-Means algorithm, while efficient, tends to produce clusters of similar size and lacks sensitivity to topographic features. on the other hand, Gaussian Mixture Models (GMM) are limited by the assumption that clusters follow Gaussian distributions. Both of these algorithms are unsuited for clustering the structures in the UMAP embeddings, since they may come in various shapes and sizes. Instead, we turn our interest to flexible, non-parametric clustering algorithms. One such algorithm is DBSCAN, which works as follows. Firstly, it selects a point and counts the number of neighbors within a radius given by the \texttt{eps} hyperparameter. Secondly, it checks whether the number of neighbors is equal to or greater than the \texttt{mean\_samples} hyperparameter. If true, the point is considered a core point, otherwise it is labeled as an outlier. Thirdly, core points that are within close proximity to each other are grouped into the same cluster. Through this method, DBSCAN is capable of capturing clusters of arbitrary shapes and sizes, mediated only by the density of points, with no prior parametric assumption. Additionally, DBSCAN does not require the user to specify the number of clusters, which is a mandatory hyperparameter in algorithms like K-Means and GMM.

    One limitation that is not addressed by DBSCAN is a measure of the robustness of its clusters. In other words, we do not know how sensitive the results are to minor variations in the hyperparameters. This problem is addressed by HDBSCAN, which distinguishes itself from DBSCAN by replacing the \texttt{eps} distance with the mutual reachability distance. Whereas \texttt{eps} is a fixed distance and thus unable to capture clusters of different densities, the mutual reachability distance varies based on the local density information and can therefore identify clusters of varying densities. Firstly, HDBSCAN calculates this mutual reachability distance between all points. Secondly, the minimum spanning tree is created, where nodes represent points, and the edges are weighted according to the mutual reachability distance. Thirdly, the algorithm tracks the evolution of the clusters as the mutual reachability distance threshold increases. The parameter $\lambda$, defined as the reciprocal of the mutual reachability distance, is evaluated at each step. Clusters begin to group together, condensing the cluster tree until only a single cluster remains, encompassing all of the data. Throughout this process, some clusters will have persisted more robustly than others. This robustness is measured by the persistence score, defined as:
\begin{align}
    \mathrm{Persistence} = \sum_{p \, \in \, \mathrm{cluster}} \left( \lambda_p - \lambda_\mathrm{birth} \right) \, ,
    \label{eqn:persistence_score}
\end{align}
    where $\lambda_p$ represents the value of $\lambda$ when the point is no longer connected to the cluster. This value is constrained between $\lambda_\mathrm{birth}$ and $\lambda_\mathrm{death}$, which respectively define the values of $\lambda$ when the cluster is born and dies.

    HDBSCAN automatically identifies the most persistent clusters, although these clusters also depend on HDBSCAN's own hyperparameters, namely \texttt{min\_cluster\_size} and \texttt{min\_samples}. To objectively select the optimal clusters, we conduct a hyperparameter optimization routine using the \texttt{Python} package \texttt{optuna}. Our loss function is the persistence score provided by the \texttt{hdbscan} package.
    
    To mitigate the risk of clustering random noise, we must define a baseline persistence score to differentiate between statistically meaningful clusters and those that are not. We generated 1000 random 5-dimensional input spaces, each containing only noise, and computed their latent space embeddings. We then ran the HDBSCAN hyperparameter optimization procedure to obtain the optimal clusters. This gave us a distribution of persistence scores, and we define the baseline score as the 99.7th percentile (3$\sigma$ equivalent) of this distribution. If the persistence scores from our samples exceed this baseline score, we regard the clusters as statistically meaningful; otherwise, we disregard them indistinguishable from random noise.

\subsection{The VM criterion}

\begin{figure}[tb!]
    \centering
    \includegraphics[width=\columnwidth]{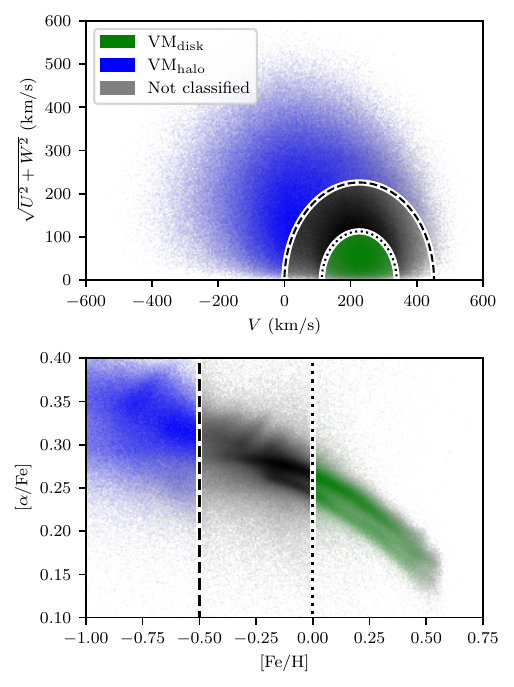}
    \caption{Kinematic and chemical parameters of the stellar particles in \texttt{m12f}. Dashed lines: $| \vec{v} - \vec{V_\mathrm{LSR}}| = |\vec{V_\mathrm{LSR}}|$ and $\mathrm{[Fe/H]} =-0.50 ~\mathrm{dex}$. Dotted lines: $| \vec{v} - \vec{V_\mathrm{LSR}}| = |\frac{1}{2} \times \vec{V_\mathrm{LSR}}|$ and $\mathrm{[Fe/H]} = 0 ~\mathrm{dex}$. Green represents the $\mathrm{VM_{disk}}$, blue represents the $\mathrm{VM_{halo}}$, and gray represents neither. \textbf{Top:} Toomre diagram. \textbf{Bottom:} [$\alpha$/Fe]--[Fe/H] chemical plane.}
    \label{fig:VM_criteria}
    \end{figure}

    In order to validate the legitimacy of the embedding components as physically meaningful stellar populations, we study their physical parameters and seek to interpret them. Since the samples only probe the stars contained within a sphere of $3 ~\mathrm{kpc}$ radius around the LSR, we expect our initial application of the embedding to distinguish between the disk and the halo, as the stars of each of these components distinguish themselves greatly in age, chemical content and kinematic profile. To achieve this, we employ a velocity and metallicity (VM) criterion, similar to the method used in \cite{Ostdiek2020A&A...636A..75O}. In our case, we want to define stars for which we have a high degree of confidence as being part of the halo, but not the disk, and vice versa. We call this $\mathrm{VM_{halo}}$ and $\mathrm{VM_{disk}}$. We know that the azimuthal velocity for the LSR of \texttt{m12f-lsr0}, a disk star, is $\vec{V_\mathrm{LSR}} = 226 ~\mathrm{km\,s^{-1}}$, on par with the bulk of the stars in sample A. Let $\vec{v}$ represent the 3D cylindrical velocity vector. Stars with $| \vec{v} - \vec{V_\mathrm{LSR}}| > |\vec{V_\mathrm{LSR}}|$ and $\mathrm{[Fe/H]} <-0.50 ~\mathrm{dex}$ are defined as being part of the $\mathrm{VM_{halo}}$, whereas stars with $| \vec{v} - \vec{V_\mathrm{LSR}}| < |\frac{1}{2} \times \vec{V_\mathrm{LSR}}|$ and $\mathrm{[Fe/H]} > 0 ~\mathrm{dex}$ are defined as being part of the $\mathrm{VM_{disk}}$, with everything in between being considered part of a transition zone that can include both halo and disk stars (see Fig. \ref{fig:VM_criteria}). These values were chosen based on an inspection of the physical parameters of the \texttt{m12f} galaxy. Figure \href{https://doi.org/10.5281/zenodo.14749285}{A.3} shows the effects of this selection on the geometry of the corresponding stellar populations. This criterion is not intended to strictly define the halo nor the disk, but rather to distinguish two highly disparate populations that are strongly associated to the halo and the disk, while minimizing contamination.
    
\begin{figure}[tb!]
    \centering
    \includegraphics[width=\columnwidth]{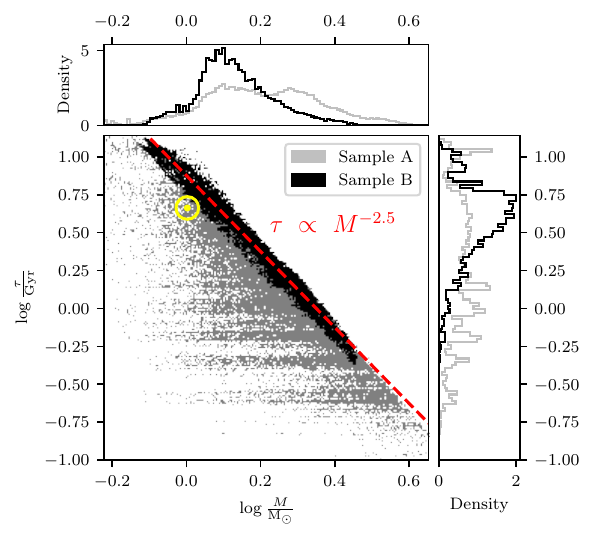}
    \caption{Age-mass distribution for sample A (silver) and sample B (black), showing the tight relationship for the seismic red-giant stars (sample B). The location of the Sun is identified by its usual symbol ($\odot$), assuming an age of 4.6 Gyr. The red dashed line follows the $\tau~\propto~M^{1 - \eta}$ scaling, where $\eta = 3.5$ \citep[cf.][]{Kippenhahn2013sse..book.....K,Davies2016AN....337..774D}.}
    \label{fig:age_mass_relationship}
    \end{figure}

    We note that stars not classified as part of the $\mathrm{VM_{disk}}$ nor the $\mathrm{VM_{halo}}$ can still overlap with these components. For instance, they could assume disk-like kinematics but low metallicity. These ``unclassified'' stars thus span from halo stars to disk stars, and everything in between, including the parameter space in which both components of the [$\alpha$/Fe] dichotomy begin to take form. This selection ensures that the $\mathrm{VM_{disk}}$ and $\mathrm{VM_{halo}}$ are adequately separated in both chemo-kinematic space and the galaxy's evolution, with the median age for the $\mathrm{VM_{disk}}$ being $2.4 ~\mathrm{Gyr}$, whereas the median age for the $\mathrm{VM_{halo}}$ is $10.4 ~\mathrm{Gyr}$. The percentage of stars contained in $\mathrm{VM_{disk}}$, $\mathrm{VM_{halo}}$ and $\mathrm{VM_{Not~Classified}}$ for samples A, B, C and D is given in Table \ref{tab:VM_fractions}. We note that, by design, only a minority of the stars are in either the $\mathrm{VM_{disk}}$ or $\mathrm{VM_{halo}}$. We also note that samples B, C, and D are undersampled in both the $\mathrm{VM_{disk}}$ and $\mathrm{VM_{halo}}$ relative to sample A, which could affect the ability to resolve both the disk and the halo. This effect originates from the selection function applied to the seismic red-giant stars. The inferred age of red-giant stars is primarily determined by the time spent in the main sequence, which follows a scaling relation of $\tau \propto M^{1 - \eta}$, where $\eta$ is the exponent in the $L$--$M$ relation for main-sequence stars \citep[][]{Kippenhahn2013sse..book.....K,Davies2016AN....337..774D}. This results in a tight age-initial mass relation for red-giant stars, which will affect the age distribution of our sample. This effect is evident in Fig.~\ref{fig:age_mass_relationship}, which compares the age and mass distributions of samples A and B. Sample A contains main-sequence stars that span a wide variety of masses and ages, whereas sample B (and, by extension, samples C and D) is constrained to masses $0.6 < M < 2.9 ~\mathrm{M_\odot}$ with a peak at $M \approx 1.2 ~\mathrm{M_\odot}$, resulting in a significant undersampling of stars younger than $\tau \approx 3 ~\mathrm{Gyr}$.

    \setlength{\tabcolsep}{1.0em} 
    \renewcommand{\arraystretch}{1.4}
    \begin{table}
    	\caption{Fraction of stars per $\mathrm{VM}$ component in each sample.}
        \label{tab:VM_fractions}
    	\centering
    	\begin{tabular}{l|lll}
    		\hline
    		\hline
    Sample      &	$\mathrm{VM_{disk}}$  &   $\mathrm{VM_{halo}}$ &   $\mathrm{VM_{Not~Classified}}$  \\
                &   (\%)                  &   (\%)                 &   (\%)                        \\
    \hline \hline
    A           &    24.4     &    6.3   &   69.3  \\
    B           &    15.1     &    3.9   &   81.0  \\
    C           &    15.2     &    3.9   &   80.9  \\
    D           &    15.2     &    3.9   &   81.0  \\
    		\hline
    	\end{tabular}
    \end{table}

\FloatBarrier
\section{Results}
\label{sec:results}
    In this section, we perform two consecutive applications of UMAP on samples A, B, C and D. The first application of UMAP, henceforth referred to as the first run, generates the latent space embedding for the entire sample. If physically meaningful components are found within these embeddings, then we perform another application of UMAP, this time on each of the components, in an attempt to subdivide them into smaller constituent subcomponents. This is referred to as the second run.

\subsection{First run}
\begin{figure*}
    \centering
    \includegraphics[width=\textwidth]{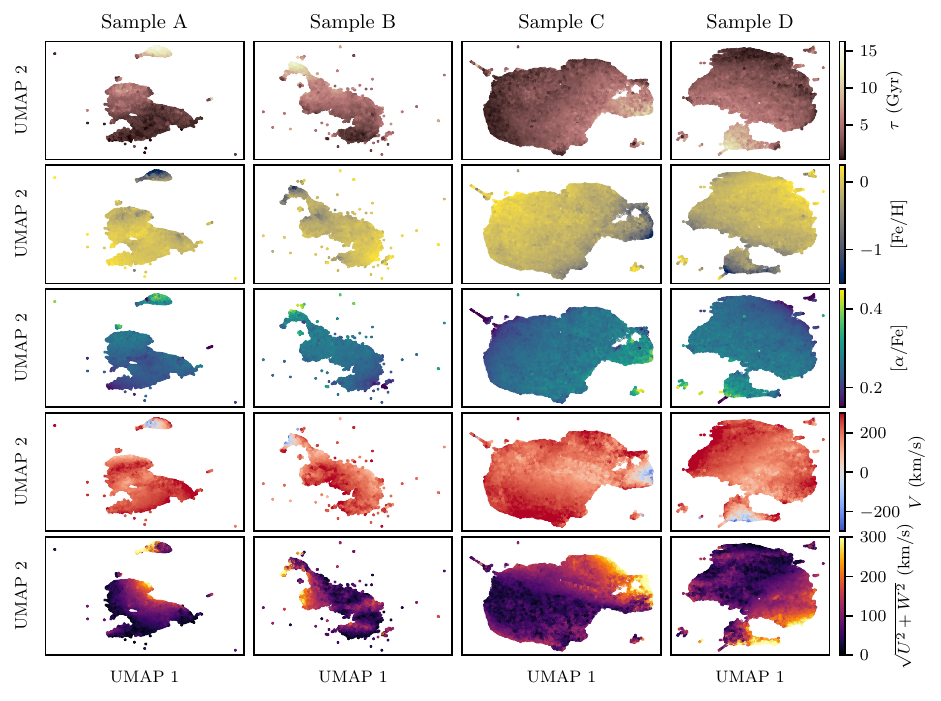}
    \caption{First run, 2D UMAP latent space embeddings for samples A, B, C and D, color-coded according to the input parameters. The specific orientation of the embeddings is not important, since there is no physical meaning to the coordinates of the latent space. What matters is the relative position of the points with respect to each other. As a result, the embeddings may appear flipped from one sample to another.}
    \label{fig:UMAP_first_run_parameters}
    \end{figure*}

    We applied the UMAP algorithm to samples A, B, C, and D, and for each sample we obtained 2D latent space embeddings. Figure \ref{fig:UMAP_first_run_parameters} illustrates these embeddings, color-coded according to the distributions of their input parameters. These parameters do not follow a strictly monotonic gradient, especially for the kinematic parameters. This information is encoded in the assumed topology of the embeddings. We notice that old, low-metallicity stars tend to agglomerate together and self-segregate from the remaining stars. Despite this, there is a clear diversity in the chemical abundances and kinematics, being the sole region where we identify retrograde stars ($V < 0 \, \mathrm{km\,s^{-1}}$). To determine the validity of these embeddings, we compute their trustworthiness scores, $T(k)$, with $k = \texttt{n\_neighbors}$ (see Sect. \ref{sec:methods} for details). We find that the trustworthiness scores for samples A, B, C, and D are $0.988$, $0.994$, $0.969$, and $0.978$, respectively. Given that the embeddings satisfy $T(k = \texttt{n\_neighbors}) > 0.90$ in all cases, we consider these embeddings to be informative and proceed with our analysis.

\begin{figure*}
    \centering
    \includegraphics[width=\textwidth]{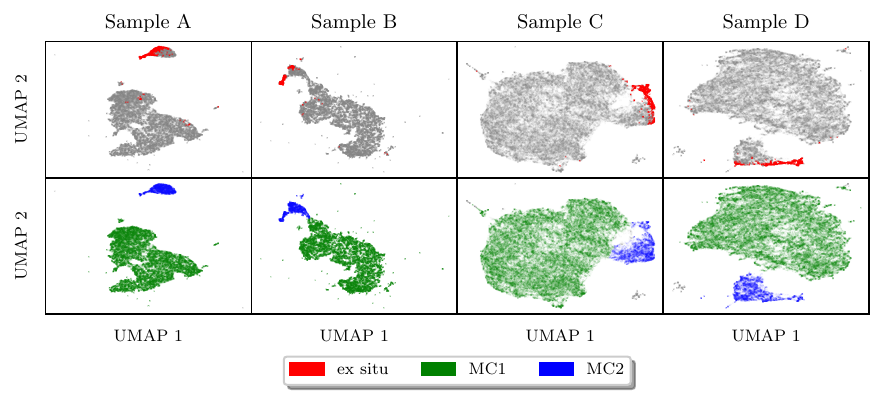}
    \caption{\textbf{Top:} First run, 2D UMAP latent space embeddings for samples A, B, C and D. Red dots correspond to ex situ stars ($d_\mathrm{form} > 30 \, \mathrm{kpc}$). \textbf{Bottom:} Selection of the major components 1 and 2 (MC1 and MC2) in green and blue, respectively.}
    \label{fig:UMAP_first_run}
    \end{figure*}

    Figure \ref{fig:UMAP_first_run} shows how there is a pronounced preference for ex situ stars (shown in red) to agglomerate within the aforementioned structure. This structure is fully disjoint for sample A. While not the case for samples B, C and D, there are nonetheless indications for the existence of a non-unified structure that appears to assume at least two distinct groups. We applied the HDBSCAN algorithm on these embeddings with a hyperparameter optimization routine, aiming to maximize the persistence score. We obtained two major components for all samples, denoted as the major component 1 (MC1) and the major component 2 (MC2). MC1 corresponds to the major component with the highest number of stars, and MC2 to the one with the lowest, which we color in green and blue, respectively. It is also worth noting that the vast majority of the accreted stars (defined as stars with $d_\mathrm{form} > 30 ~\mathrm{kpc}$) are located in MC2, with $97.6\%$ to $98.7\%$ of them found in this component, depending on the sample. At no point was $d_\mathrm{form}$, the parameter used to define ex situ stars, included in the generation of the embeddings. The fraction of stars in MC2 classified as accreted ranges from $10.9\%$ to $16.6\%$, depending on the sample.

    The next step is to determine if the major components MC1 and MC2 correspond to physically meaningful stellar populations. In particular, we aim to verify if our assumption of disk and halo segregation is correct. If this assumption is correct, we would expect that MC1 and MC2 contain the disk and the halo separately, as defined by the VM criterion, i.e., $\mathrm{VM_{disk}}$ and $\mathrm{VM_{halo}}$ (see Sect.~\ref{sec:methods}). To assess this, we construct a $3 \times 3$ confusion matrix, with truth values defined as $\mathrm{VM_{disk}}$, $\mathrm{VM_{halo}}$ and $\mathrm{VM_{Not~Classified}}$. The predicted values are defined as MC1, MC2 and ``other'', where ``other'' refers to outlier stars that were not assigned to any cluster by HDBSCAN. Since there is no predicted class analog for the truth class $\mathrm{VM_{Not~Classified}}$, we exclude this from our evaluation metrics. We do include, however, the predicted class "other". For example, stars with truth value $\mathrm{VM_{disk}}$ assigned a prediction label othar than MC1 (i.e., MC2 or ``other'') are defined as false negatives. The reverse situation is also true. Due to the size discrepancy between MC1 and MC2, the predictive power of MC1 dominates and skews the results. To mitigate this, we calculate the specificity and sensitivity, which, for a given truth class $C_i$ (either $\mathrm{VM_{disk}}$ or $\mathrm{VM_{halo}}$), are defined as:
    \begin{align}
        \mathrm{Sensitivity}_{C_i} &= 
        \frac{\mathrm{TP}_{C_i}}{\mathrm{TP}_{C_i} + \mathrm{FN}_{C_i}} \, ,
        \\
        \mathrm{Specificity}_{C_i} &= 
        \frac{\mathrm{TN}_{C_i}}{\mathrm{TN}_{C_i} + \mathrm{FP}_{C_i}} \, ,
    \end{align}
    where TP, TN, FP and FN correspond to true positive, true negative, false positive and false negative, respectively. This allows us calculate the informedness score for each truth class:
    \begin{align}
        \mathrm{Informedness}_{C_i} &= \mathrm{Specificity}_{C_i} + \mathrm{Sensitivity}_{C_i} - 1 \, ,
    \end{align}
    which can then be combined in an arithmetic mean to provide the total informedness. The total informedness ranges from $-1$ to $1$, with values close to $1$ indicating a good classification, values close to $-1$ suggesting that the labels have been inverted, and values close to $0$ indicating a poor classification.

    We find the informedness scores for samples A, B, C ,and D to be, respectively, $0.989$, $0.966$, $0.966$, and $0.973$. These values indicate that MC1 and MC2 contain, respectively, the disk and the halo, with little cross-contamination. Our results are summarized in Table \ref{tab:validation_metrics}.

    \setlength{\tabcolsep}{1.0em} 
    \renewcommand{\arraystretch}{1.4}
    \begin{table}
    	\caption{Scoring metrics for samples A, B, C and D.}
        \label{tab:validation_metrics}
    	\centering
    	\begin{tabular}{l|lllll}
    		\hline
    		\hline
    Sample      &	$T(k=\mathrm{n\_neighbors})$  &   Informedness  \\
                &                                 &        \\
    \hline \hline
    A           &    0.988     &    0.989  \\
    B           &    0.994     &    0.966   \\
    C           &    0.969     &    0.966  \\
    D           &    0.978     &    0.973    \\
    		\hline
    	\end{tabular}
        \tablefoot{$T(k)$ is the trustworthiness score for $k$ nearest neighbors. The informedness score is the arithmetic mean of the informedness scores for the disk and halo.}
    \end{table}

\subsection{Second run}
\begin{figure*}
    \centering
    \includegraphics[width=\textwidth]{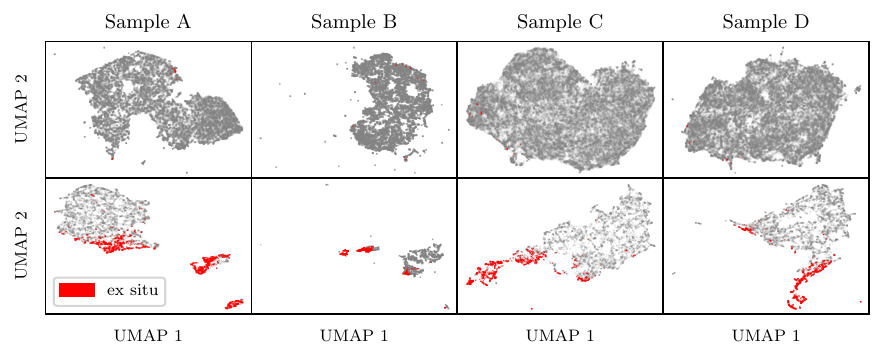}
    \caption{2D UMAP latent space embeddings for the major components MC1 (top row) and MC2 (bottom row) in each of the samples. Red indicates ex situ stars.}
    \label{fig:UMAP_second_run_exsitu}
    \end{figure*}

    A closer inspection of the latent space embeddings in Fig. \ref{fig:UMAP_first_run} reveals further topographic features within MC1 and MC2. For samples A and B, MC1 appears to be either trying to dislodge or elongate itself, which is not the case for samples C and D. In all samples, we also notice that MC2 possesses a tail. These features hint at the existence of a more nuanced structure. Motivated by this, we performed a second run for MC1 and MC2, taken separately, for each of the samples. We ran UMAP 100 times, applying an HDBSCAN optimization routine to each embedding, and kept track of the persistence score and the number of clusters found. We retain the embeddings with the mode number of clusters and, from that pool, focus our attention on the embedding for which HDBSCAN returned the highest persistence score. These final embeddings, obtained for MC1 and MC2 in each of the samples, are displayed in Fig.~\ref{fig:UMAP_second_run_exsitu}. We note once again the tendency for ex situ stars to agglomerate together in the MC2 embeddings, potentially forming predominantly ex situ substructures. We show the corresponding distributions of the input parameters in Figs.~\href{https://doi.org/10.5281/zenodo.14749285}{C.1} and \href{https://doi.org/10.5281/zenodo.14749285}{C.2} of the Appendix.

\begin{figure}
    \centering
    \includegraphics[width=\columnwidth]{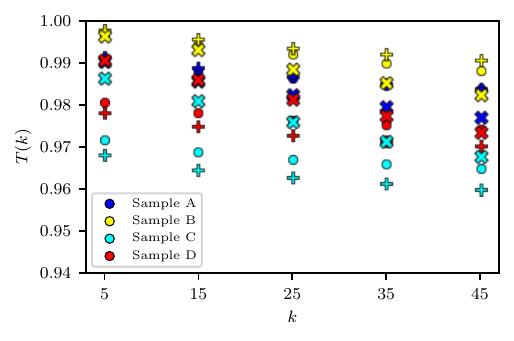}
    \caption{Trustworthiness score, $T$, as a function of the number of nearest neighbors, $k$, for the UMAP embeddings obtained from samples A, B, C and D. Circles represent the first run embeddings, plus signs the second run MC1 embeddings and crosses the second run MC2 embeddings. All values satisfy $T(k) > 0.90$.}
    \label{fig:trustworthiness_score_all}
    \end{figure}

    Similar to the first run, we compute the trustworthiness score $T(k = \texttt{n\_neighbors})$ for the second run MC1 and MC2 embeddings and compare them to the trustworthiness score of the first run embeddings. To assess the sensitivity of $T$ as a function of $k$, we also compute the trustworthiness score in the vicinity of $k = \texttt{n\_neighbors}$. These results are plotted in Fig.~\ref{fig:trustworthiness_score_all}, displaying a small downward trend as $k$ is increased, but nonetheless satisfying $T(k) > 0.90$ for all scenarios.

\begin{figure*}
    \centering
    \includegraphics[width=\textwidth]{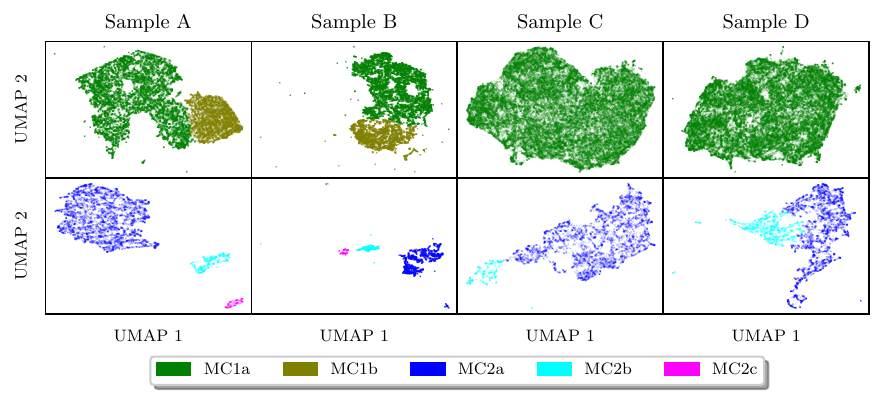}
    \caption{2D UMAP latent space embeddings for the major components MC1 (top row) and MC2 (bottom row) in each of the samples, color-coded according to the identified subcomponents. \textbf{Top:} MC1, subdivided into MC1a (green) and MC1b (olive). \textbf{Bottom:} MC2, subdivided into MC2a (blue), MC2b (cyan) and MC2c (magenta).}
    \label{fig:UMAP_second_run_all}
    \end{figure*}

    The HDBSCAN clustering of the MC1 and MC2 embeddings is shown in Fig.~\ref{fig:UMAP_second_run_all}. While the algorithm initially converged to two clusters for MC1 from samples C and D, the persistence score did not surpass the baseline required to confidently assert them as statistically meaningful and were thus merged into one. This left us with two MC1 subcomponents for samples A and B, dubbed MC1a and MC1b, and just one for samples C and D, which we refer to as MC1a for consistency. As for MC2, HDBSCAN converged to three clusters for samples A and B, which we dub MC2a, MC2b, and MC2c. For samples C and D, we obtain only two subcomponents, which we name MC2a and MC2b for consistency. These subcomponents are named based on the number of stars within them, and do not necessarily refer to the same population across samples.

\subsection{Interpreting the subcomponents}
    Whether the subcomponents found by HDBSCAN, applied to the UMAP latent space embeddings, correspond to genuinely distinct stellar populations remains to be seen. This would be straightforward if the stars had their population membership determined a priori. However, much like in a real scenario, we lack such direct labeling and must instead rely on inferences based on stellar parameters. Since these samples are based on synthetic data, we also have access to meta-parameters that can be used to validate our findings. One such useful meta-parameter is the distance of formation, $d_\mathrm{form}$, which allows us to get a sense of the star formation history (SFH) and the merger history of the synthetic \texttt{m12f} galaxy (see Fig.~\ref{fig:m12f_dform}). If our findings are meaningful, they should not only be sensibly segregated in the 5D input parameter space, but also probe different evolutionary phases of the galaxy, as revealed by the simulation's meta-parameters.

\begin{figure*}
    \centering
    \includegraphics[width=\textwidth]{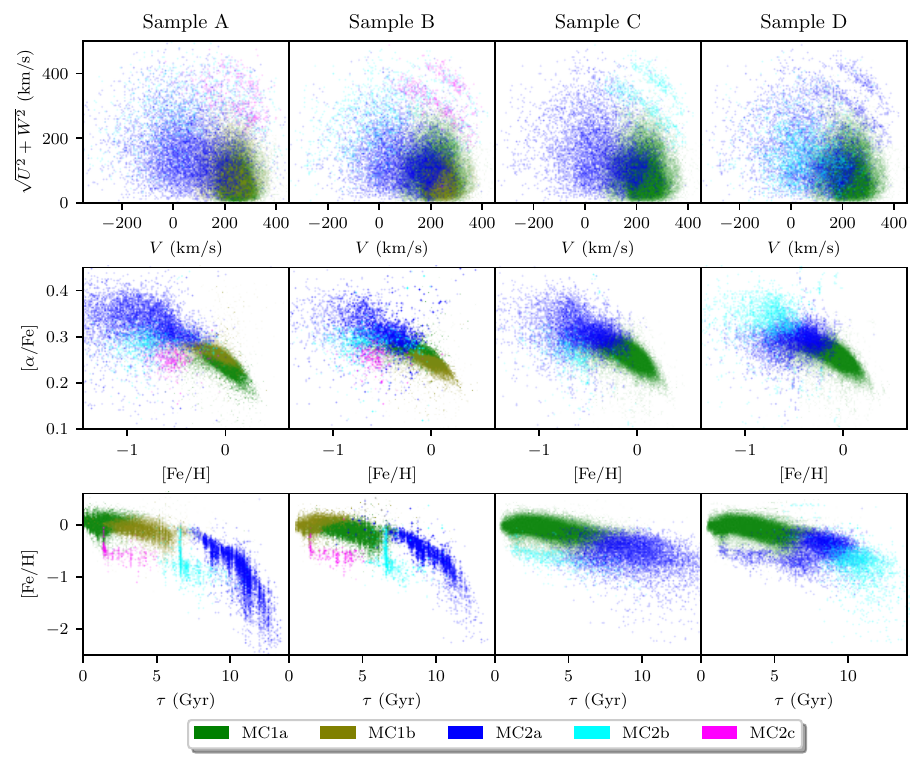}
    \caption{Distributions of the input parameters for samples A, B, C and D, color-coded according to the second run subcomponents. \textbf{Top:} Toomre diagram. \textbf{Middle:} [$\alpha$/Fe] versus [Fe/H]. \textbf{Bottom:} Stellar ages versus [Fe/H].}
    \label{fig:input_space_second_run_all}
    \end{figure*}

    We begin with an analysis of the input parameters. Figure \ref{fig:input_space_second_run_all} shows the distributions of the input parameters (Toomre diagram, [$\alpha$/Fe] vs [Fe/H], and stellar ages vs [Fe/H]) for all second run subcomponents of samples A, B, C and D. For a more detailed view, we provide Figs. \href{https://doi.org/10.5281/zenodo.14749285}{C.3}, \href{https://doi.org/10.5281/zenodo.14749285}{C.4}, and \href{https://doi.org/10.5281/zenodo.14749285}{C.5} in the Appendix, respectively comparing the kinematic, chemical, and stellar age parameters for each sample. We note the tendency for the MC1 subcomponents to assume a narrow prograde distribution, in contrast to the wider kinematic distributions of the MC2 subcomponents, which include all retrograde stars. In chemical space, MC1 subcomponents assume a tight [$\alpha$/Fe]--[Fe/H] correlation at the high-metallicity end, whereas the MC2 subcomponents span a wide range in [$\alpha$/Fe] at low metallicity. The $\tau$--[Fe/H] distributions reveal what appear to be two distinct, parallel chemical evolutionary tracks, with the MC1 subcomponents consistently occupying the younger portion of the higher-[Fe/H] track. These features are consistent with a distinction between the disk and halo as observed in the Milky Way, where the disk is comprised of young, metal-rich stars with a tight velocity dispersion around $\vec{v} \approx \vec{v}_\mathrm{LSR}$, contrasting with the old, metal-poor stars of the halo, which are further characterized by a large velocity dispersion that encompasses stars with retrograde orbits \citep{Helmi2008A&ARv..15..145H, Bland-Hawthorn2016ARA&A..54..529B, GaiaDR2DiscKinematics2018}. The nuances then arise in the interpretation of each subcomponent. Since we have different results for each sample, we will go through them one by one.

    For sample A, the subcomponents MC1a and MC1b appear to correspond to two chemically distinct disk populations, in line with the chemical dichotomy shown in Fig.~\ref{fig:m12f_chemical_evolution}. MC1a is young and occupies the lower-[$\alpha$/Fe] track, in contrast to MC1b's older age and higher [$\alpha$/Fe]. As for MC2, the subcomponent MC2a is the most populous, displaying a large velocity dispersion and the highest [$\alpha$/Fe], and it further contains the oldest stars in the sample. In contrast, MC2b and MC2c both display predominantly large $\sqrt{U^2 + W^2}$ velocities, with MC2c in particular being entirely prograde. MC2b and MC2c also display lower $\mathrm{[\alpha/Fe]}$ compared to MC2a, while following a lower-$\mathrm{[Fe/H]}$ chemical evolutionary track. Features in the [$\alpha$/Fe]--[Fe/H] plane are explained by differences in the rates and yields of the dominant nucleosynthesis processes, including type Ia SNe, CC-SNe, and stellar winds from asymptotic giant-branch stars. These, in turn, are dependent on the mass of the progenitor dark-matter halo, which influences star formation rates. The trends found here for subcomponents MC2b and MC2c, compared to MC2a, are reminiscent of the observed trends resulting from the \textit{Gaia}-Enceladus merger \citep{Helmi2018Natur.563...85H}. Indeed, the chemical evolutionary history of \texttt{m12f} appears to be explained by the accretion of at least one lower-mass galaxy \citep[see, e.g.,][]{Vincenzo2019MNRAS.487L..47V}. The combined chrono-chemo-kinematic analysis strongly suggests that MC2a represents the primordial stellar halo, generated in situ, whereas the subcomponents MC2b and MC2c represent accreted populations.
    
    For sample B, the MC2a, MC2b, and MC2c subcomponents are identical to those found for sample A. Whereas for sample A, MC1a and MC1b respectively corresponded to the young and old disk populations, the reverse\footnote{The labeling is ordered as a function of the number of stars in each subcomponent. For samples A and B, the subcomponents MC1a and MC1b have their labels switched, but otherwise correspond to the same populations.} is found for sample B. These subcomponents appear to represent the high-$\alpha$ and low-$\alpha$ disks, similar to the MC1 subcomponents found for sample A. However, the partitioning of the disk was not possible for samples C and D. This is explained by the fact that samples C and D contain perturbed input parameters, leading to an inability to differentiate between the high-$\alpha$ and low-$\alpha$ disk subcomponents. As a result, the ability to distinguish the MC2 subcomponents for samples C and D is also affected. For sample C, the subcomponent MC2a appears to encompass the in situ halo and a significant portion of the accreted stars, while MC2b encompasses the remaining accreted stars. For sample D, we find that MC2a captures all accreted stars as well as the younger portion of the in situ halo, while MC2b captures the older portion of the halo.

\begin{figure*}
    \centering
    \includegraphics[width=\textwidth]{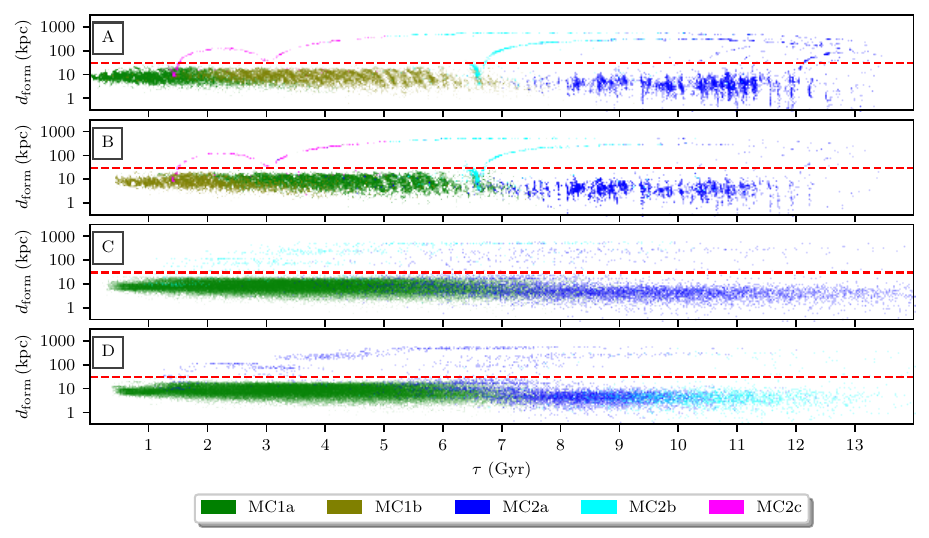}
    \caption{Distance of formation ($d_\mathrm{form}$) as a function of age for samples A, B, C and D, color-coded according to the second run subcomponents. The horizontal dashed line marks $d_\mathrm{form} = 30~\mathrm{kpc}$.}
    \label{fig:dform_subcomponents}
    \end{figure*}

\begin{figure*}
    \centering
    \includegraphics[width=\textwidth]{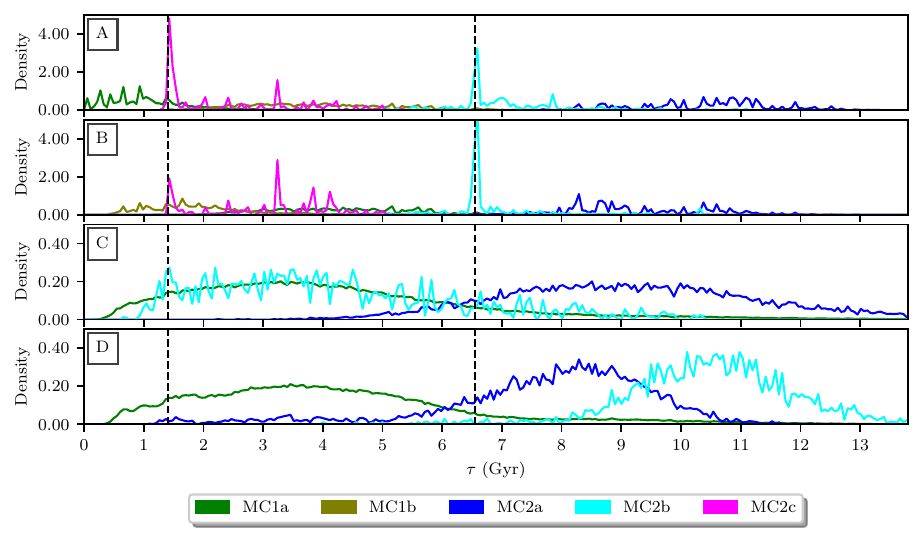}
    \caption{Kernel density estimation of the stellar ages for samples A, B, C and D, color-coded according to the second run subcomponents. Vertical dashed lines at $\tau = 1.4$ and $6.5 ~\mathrm{Gyr}$ mark the occurrence of each major accretion event.}
    \label{fig:age_histograms_subcomponents}
    \end{figure*}

    To provide a fuller picture, we plot $d_\mathrm{form}$ as a function of age in Fig.~\ref{fig:dform_subcomponents}. Let us begin by considering sample A. Focusing on the accreted stars ($d_\mathrm{form} > 30 ~\mathrm{kpc}$), we observe that the subcomponents MC2b and MC2c trace the two major accretion events (cf.~Fig.~\ref{fig:m12f_dform}), although with some degree of contamination. Shifting our attention to the in situ stars ($d_\mathrm{form} < 30 ~\mathrm{kpc}$), we identify the bulk of MC2a, which we associate with the primordial stellar halo. Leftward of the 8 Gyr mark, we notice the first stars cataloged as belonging to MC1b, with the other disk component (MC1a) emerging only in the past few Gyr. Interestingly, the two major accretion events traced by MC2b and MC2c are associated with bursts of in situ star formation, at $\tau = 6.5$ and $1.4 ~\mathrm{Gyr}$, respectively (see Fig.~\ref{fig:age_histograms_subcomponents}). The origin of the low-$\alpha$ disk subcomponent, identified as MC1a, can be explained by the inflow of metal-poor gas brought into the disk by the second major accretion event (MC2c), which is mostly characterized by ages younger than $\tau \approx 1.4 ~\mathrm{Gyr}$.

    The subcomponents found for sample B are largely analogous to the ones found for sample A, with MC2a, MC2b and MC2c representing the exact same populations. For MC1a and MC1b, however, the situation is not as straightforward. Firstly, the labels are switched compared to sample A, since now the older subcomponent (MC1a) contains more stars. Secondly, as shown in Fig.~\ref{fig:age_histograms_subcomponents}, the bulk of the younger subcomponent (MC1b) contains stars older than the time of the second major accretion. This means that, although this subcomponent contains the low-$\alpha$ disk, it also contains a portion of the high-$\alpha$ disk. Thus, there is an added difficulty in adequately segregating the two disk subcomponents, likely due to the undersampling of very young stars relative to sample A, which, in turn, is rooted in the selection of seismic red giants when building sample B (see Fig.~\ref{fig:age_mass_relationship}). For sample C, there is no segregation of MC1 into subcomponents, whereas MC2 is segregated into two, rather than three, subcomponents. The subcomponent MC2a captures the in situ halo as well as the stars from the first accreted satellite galaxy, with MC2b representing the stars belonging to the second accreted galaxy. For sample D, MC1 is also not segregated into subcomponents, and the two subcomponents found for MC2 differ from those found in sample C. MC2a now corresponds to the full accreted population as well as the younger portion of the in situ halo, while MC2b corresponds to the older portion of the in situ halo. The detection of fewer subcomponents for samples C and D is explained by their perturbed input parameters, which blur the fine details that could have been used to analyze these subcomponents. The segregation of the low-$\alpha$ and high-$\alpha$ disks then becomes unfeasible because their separation in [$\alpha$/Fe] is of the order of the adopted uncertainties. The perturbations to the chemical parameters, along with the perturbations to the stellar age, also hinder the ability to adequately segregate the halo subpopulations. These effects are evident when comparing sample B with samples C and D, as shown in Fig.~\ref{fig:input_space_second_run_all}. Our interpretation of each identified subcomponent across the four samples is provided in Table \ref{tab:subcomponents_interpretation}.

    \setlength{\tabcolsep}{1.0em} 
    \renewcommand{\arraystretch}{1.4}
    \begin{table*}
    	\caption{Interpretation of each identified subcomponent across samples A, B, C and D.}
        \label{tab:subcomponents_interpretation}
    	\centering
    	\begin{tabular}{l|llll}
    		\hline
    		\hline
                     &	Sample A   &   Sample B    &   Sample C    &   Sample D    \\
    \hline \hline
    MC1a             &    Low-$\alpha$ disk     &   High-$\alpha$ disk  &   N/A   & N/A   \\
    MC1b             &    High-$\alpha$ disk    &   Low-$\alpha$ disk   &   N/A   & N/A   \\
    MC2a             &    in situ halo    &   in situ halo   &   in situ halo + first merger   & Mergers + younger in situ halo   \\
    MC2b             &    First merger    &   First merger   &   Second merger   & Older in situ halo   \\
    MC2c             &    Second merger    &   Second merger   &   N/A   & N/A   \\
    \hline
    	\end{tabular}
    \end{table*}

\section{Summary and conclusions}
\label{sec:summary_conclusions}
    In this work, we used a cosmological zoom-in galaxy for with a known evolutionary history and generated four synthetic stellar samples (A, B, C, and D): sample A consists of 75,000 randomly selected stars in all evolutionary states, sample B consists of a TESS-like seismic red-giant selection containing $76,168$ stars, and samples C and D are perturbed versions of sample B. Considering a 5D input parameter space that includes stellar ages, kinematics ($V$ and $\sqrt{U^2 + W^2}$) and chemical abundances ($\mathrm{[Fe/H]}$ and $\mathrm{[\alpha/Fe]}$), we applied the unsupervised manifold learning algorithm UMAP to generate latent space embeddings. The HDBSCAN clustering algorithm was then applied multiple times to obtain the hyperparameters that maximized the persistence score, thereby retrieving the most persistent clusters. All embeddings were validated based on the trustworthiness score, $T(k)$, which consistently returned values satisfying $T(k) > 0.90$. Our findings are as follows:

    \begin{enumerate}
        \item A first run of the method demonstrated a robust separation of the disk and halo for all samples (see Fig.~\ref{fig:UMAP_first_run}), confirmed by our test of the VM criterion, which returned informedness values greater than $0.95$ (see Table \ref{tab:validation_metrics}). This shows that the inclusion of realistic uncertainties on the input parameters (i.e., sample C) allows the segregation of these two components using a purely automated and unsupervised approach. Moreover, we demonstrate that meaningful information can be inferred from the topology of the embeddings, even in the absence of purely fully structures.

        \item We subsequently conducted a second run of the method on both the disk-like and halo-like components, identifying five subcomponents for the pristine samples A and B and three subcomponents for the perturbed samples C and D. For samples A and B, we were able to identify the high-$\alpha$ and low-$\alpha$ disks. The latter was generated from the inflow of chemically impoverished gas associated with the second major accretion event ($\tau \approx 1.4 ~\mathrm{Gyr}$; see Figs.~\ref{fig:m12f_chemical_evolution} and \ref{fig:m12f_dform}). Since this population was formed relatively recently, an undersampling of the youngest stars affects the ability to resolve it. This effect is shown in Fig.~\ref{fig:age_histograms_subcomponents}, where the low-$\alpha$ disk is still able to be identified for sample B (MC1b), but at the expense of accuracy, being grouped together with in situ stars that precede the accretion event at $\tau \approx 1.40 ~\mathrm{Gyr}$. In the case of samples C and D, we were unable to differentiate between the two disk components. We attribute this to the very tight [$\alpha$/Fe] bimodality in the synthetic galaxy, which is eventually blurred as a result of the perturbation to the chemical abundances.

        \item The halo-like component of samples A and B was segregated into three subcomponents: the in situ halo (MC2a), the first accreted population (MC2b), and the second accreted population (MC2c). For sample C, the second accreted population was segregated from the remainder of the halo. This loss in resolution is attributed to the perturbation of the input parameters. Similar or better results were expected for sample D, as it was perturbed using more optimistic uncertainties. However, sample D segregated the oldest halo stars ($\tau > 10 ~\mathrm{Gyr}$) from the remainder of the halo, which includes both accreted populations and a sizable portion of the in situ halo.
    \end{enumerate}
    
    We have shown that the application of the unsupervised manifold learning algorithm UMAP, along with an automated HDBSCAN optimization routine, can provide key inferences on the nature and number of stellar populations without any such prior assumptions. This was achieved by intelligently projecting the high-dimensional input data into a lower-dimension, latent space embedding. Various validation procedures show this to be a valid method to reliably and accurately depict the underlying non-linear and non-parametric structure hidden in the high-dimensional data. We have also shown that, by segregating the stars according to the embedded manifold structure, we can divide disk-like from halo-like structures for all four of our samples, including a TESS-like sky coverage of a red-giant sample with realistic \textit{Gaia} DR3 kinematic, APOGEE DR17 chemical and seismic age uncertainties. We compared UMAP to other algorithms, highlighting the limitations of PCA, and showed that UMAP is capable of competing with, and even surpassing, t-SNE in terms of computational efficiency and its ability to discern stellar populations. These comparisons are provided in Appendix \href{https://doi.org/10.5281/zenodo.14749285}{D}.
    
    We have also shown that, when adopting unperturbed (i.e., pristine) input parameters, the method can segregate subcomponents belonging to the disk and the halo. Although these results are not replicated for the perturbed samples, we emphasize that we made use of only five stellar parameters, these being $\{V, UW, \mathrm{[Fe/H]}, \mathrm{[\alpha/Fe]}, \tau \}$. However, we identify potential in the method to discern additional stellar populations, such as in situ vs ex situ stars, when dealing with realistic data sets. UMAP is expected to provide deeper insights into stellar populations through the inclusion of additional stellar parameters. Examples include angular momentum, potential energy, integrals of motion, orbital eccentricity, and other chemical abundance ratios, such as s-process elements ejected by the stellar winds of asymptotic giant-branch stars and r-process elements produced in neutron star mergers.

\section*{Data availability}
The Appendix is provided through the following URL: \href{https://doi.org/10.5281/zenodo.14749285}{https://doi.org/10.5281/zenodo.14749285}.

\begin{acknowledgements}
This work was supported by Fundação para a Ciência e a Tecnologia (FCT) through the research grants UIDB/04434/2020 (DOI: 10.54499/UIDB/04434/2020) and UIDP/04434/2020 (DOI: 10.54499/UIDP/04434/2020).
T.L.C.~is supported by FCT in the form of a work contract (CEECIND/00476/2018).
D.B.~acknowledges funding support by the Italian Ministerial Grant PRIN 2022, ``Radiative opacities for astrophysical applications'', no.~2022NEXMP8, CUP C53D23001220006. A.M.~acknowledges support from the European Research Council Consolidator Grant funding scheme (project ASTEROCHRONOMETRY, G.A.~no.~772293, \url{http://www.asterochronometry.eu}).
\end{acknowledgements}

\bibliographystyle{aa}
\bibliography{references}

\end{document}